\documentclass[aps,prb,twocolumn,superscriptaddress,showpacs,showkeys]{revtex4-2}

\usepackage{color}
\usepackage{graphicx}
\usepackage{dcolumn}
\usepackage{bm}
\usepackage{float}
\usepackage[mathlines]{lineno}
\usepackage{tabularx}
\usepackage[colorlinks,linkcolor=blue,anchorcolor=blue,citecolor=blue,urlcolor=blue,hyperindex,CJKbookmarks]{hyperref}
\usepackage{footnote}
\usepackage{amsmath}
\usepackage{txfonts}
\usepackage{mathptmx}
\usepackage{ragged2e}
\usepackage{booktabs,makecell, multirow, tabularx}
\usepackage{multirow}
\usepackage{braket}
\usepackage{gensymb}
\usepackage{array}
\usepackage{makecell} 
\usepackage{booktabs}

\newcolumntype{L}[1]{>{\raggedright\let\newline\\\arraybackslash\hspace{0pt}}m{#1}}
\newcolumntype{C}[1]{>{\centering\let\newline\\\arraybackslash\hspace{0pt}}m{#1}}
\newcolumntype{R}[1]{>{\raggedleft\let\newline\\\arraybackslash\hspace{0pt}}m{#1}}

\begin{document}

\title{Phase diagrams and two key factors to superconductivity of Ruddlesden-Popper nickelates}

\author{Zhenfeng Ouyang}\affiliation{School of Physics and Beijing Key Laboratory of Opto-electronic Functional Materials $\&$ Micro-nano Devices, Renmin University of China, Beijing 100872, China}\affiliation{Key Laboratory of Quantum State Construction and Manipulation (Ministry of Education), Renmin University of China, Beijing 100872, China}
\author{Rong-Qiang He}\email{rqhe@ruc.edu.cn}\affiliation{School of Physics and Beijing Key Laboratory of Opto-electronic Functional Materials $\&$ Micro-nano Devices, Renmin University of China, Beijing 100872, China}\affiliation{Key Laboratory of Quantum State Construction and Manipulation (Ministry of Education), Renmin University of China, Beijing 100872, China}
\author{Zhong-Yi Lu}\email{zlu@ruc.edu.cn}\affiliation{School of Physics and Beijing Key Laboratory of Opto-electronic Functional Materials $\&$ Micro-nano Devices, Renmin University of China, Beijing 100872, China}\affiliation{Key Laboratory of Quantum State Construction and Manipulation (Ministry of Education), Renmin University of China, Beijing 100872, China}\affiliation{Hefei National Laboratory, Hefei 230088, China}

\date{\today}

\begin{abstract}
The discovery of superconductivity in Ruddlesden-Popper (RP) nickelates has drawn great attention. Many works have been done to study the superconductivity as well as to find more superconducting RP nickelates. However, there is a lack of general understanding regarding the key factors that contribute to the superconductivity of RP nickelates. Here, we systematically study the series of RP nickelates under doping or high-pressure conditions by means of density functional theory plus dynamical mean-field theory. We find that enhanced quasi-particle weights and local spin fluctuation of the Ni-$e_g$ orbitals are commonly realized by hole doping or high pressure in the known superconducting RP nickelates, suggesting that they are crucial to the superconductivity. We also summarize experimentally synthesized RP nickelates into phase diagrams with local spin moment and local entanglement entropy as parameters, where phases of spin density wave/antiferromagnetism, superconductivity, and Fermi liquid are distinguished. At last, we predict a promising candidate for superconducting RP nickelates, which is constructed in a ``bilayer-trilayer'' stacking sequence.

\end{abstract}


\maketitle

\section{Introduction}
Extensive studies has been made since discoveries of superconductivity in various Ruddlesden-Popper (RP) nickelates, for example high-pressure bulk bilayer nickelate La$_3$Ni$_2$O$_7$ (2222-La327)~\cite{Sun-nature} and trilayer La$_4$Ni$_3$O$_{10}$ (La4310)~\cite{Zhu2024,Li2024,Li_2024,zhang2023}, etc~\cite{Ko2024,Zhou2025,shi2025,PhysRevLett.133.146002}. Although there is preliminary consensus regarding unconventional superconducting pairing in RP nickelates~\cite{Ouyang2024,You2025}, the key factors to the emergence of superconductivity remain unclear. Some issues that may be closely related to superconductivity were successively proposed, such as oxygen vacancy, structural phase transition, etc. Here, we review some of them. 

(i) Oxygen vacancies were reported in both bulk 2222-La327~\cite{Sun-nature} and La4310~\cite{Zhu2024}, which was proposed to be important to superconductivity. The spectroscopic result indeed suggested the existence of oxygen vacancies in bulk 2222-La327~\cite{Dong2024}. By replacing La with Pr, oxygen vacancy is suppressed and bulk superconductivity with large superconducting volume fractions is confirmed eventually in 2222-La327~\cite{Wang2024}. Both theoretical and experimental reports suggested that oxygen vacancy is harmful to superconductivity~\cite{PhysRevLett.131.236002,wang2025}.

(ii) Several reports claimed that there is a competing monolayer-trilayer structure (1313-La327) in La$_3$Ni$_2$O$_7$ compounds~\cite{Chen2024jacs,Wang2024ino,PhysRevLett.133.146002}, which may also be superconducting under high pressure~\cite{PhysRevLett.133.146002}. Theoretical studies suggested that the trilayer and monolayer structures in 1313-La327 show similar electronic structures and correlation to La4310 and hole-doped La$_2$NiO$_4$, respectively~\cite{PhysRevB.111.125111}. And a very recent report claimed that superconductivity was found in monolayer-bilayer La$_5$Ni$_3$O$_{11} ($1212-La5311)~\cite{shi2025}. It is expected that hybrid stacking RP nickelates~\cite{PhysRevMaterials.8.053401} will begin to attract more attentions. 

(iii) When increasing pressure, there is a structural phase transition from ambient orthorhombic phase to high-pressure tetragonal phase with the out-of-plane Ni-O-Ni angle turning from 168$^\circ$ to 180$^\circ$ in 2222-La327~\cite{Sun-nature}. A similar transition was also found in La4310~\cite{Zhu2024}. These structural phase transitions are considered to be crucial to high-pressure superconductivity in RP nickelates. However, there is no trace of superconductivity in 1313-La327~\cite{Chen2024jacs} and 1212-La5311~\cite{shi2025} with 180$^\circ$ out-of-plane Ni-O-Ni angle under ambient pressure, and even in a tetragonal phase of La4310 under both ambient and high pressures~\cite{shi2025absencesuperconductivitydensitywavetransition}. These findings suggest that there is no inevitable connection between tetragonal structure with 180$^\circ$ out-of-plane Ni-O-Ni angle and superconductivity in RP nickelates. 

(iv) The $\gamma$ pocket contributed by the Ni-3$d_{z^2}$ and O-2$p$ orbitals around the $M$ point of high-pressure 2222-La327 is considered to be important to superconductivity~\cite{Yang2024nc,PhysRevLett.131.236002,PhysRevLett.131.126001}. For example, a random-phase approximation investigation suggests that the emergent $\gamma$ pocket provides strong Fermi surfaces nestings, which plays a crucial role in superconducting pairing~\cite{PhysRevLett.131.236002}. Applying hole doping under ambient pressure may be considered an effective approach to reproduce high-pressure band structures in 2222-La327. And recent superconducting film 2222-La327 shows a trace of Sr doping under ambient pressure~\cite{li2025photoemissionevidencemultiorbitalholedoping}.

(v) It is inevitable to discuss AFM correlation in unconventional superconductivity. Different from the parent of cuprates, no long-range AFM order is observed in RP nickelates, although a common density-wave like signal is found in the measurements of resistance~\cite{Sun-nature,Zhu2024,Chen2024jacs,shi2025}. The results of resonant inelastic X-ray scattering~\cite{Chen2024rixs}, positive muon spin relaxation~\cite{PhysRevLett.132.256503}, nuclear magnetic resonance~\cite{ZHAO2025}, and neutron diffraction~\cite{plokhikh2025unravelingspindensitywave} show a trace of SDW in bulk 2222-La327 under ambient. A report regarding 1212-La5311 shows that SDW is robust against the structural transition when increasing pressure~\cite{shi2025}. And superconductivity emerges after suppressing SDW under high pressure, which strongly suggests that there is a close competition between SDW and superconductivity. However, quantitative descriptions of this competition are still lacking.

As for the issues mentioned above, oxygen vacancy is supposed to be harmful to superconductivity~\cite{PhysRevLett.131.236002,wang2025}. The discovery of superconducting high-pressure hybrid RP 1212-La5311~\cite{shi2025} shows that hybrid RP nickelates are also promising superconductor candidates. The separation between structural phase transition and emergence of superconductivity in 1212-La5311~\cite{shi2025} may suggest that superconductivity is affected by more complicated factors than structural phase transition. How correlated electronic structure and magnetism affect superconductivity in RP nickelates is still worthy of further study. The current wealth of experimental and theoretical researches provide us good opportunities to find the key factors to superconductivity of RP nickelates.

In this work, we perform investigations regarding film 2222-La327, hybrid 1212-La5311, and La$_2$NiO$_4$ by using the density functional theory plus dynamical mean-field theory (DFT+DMFT). We find that large quasi-particle weight and spin fluctuation are the two crucial factors to superconductivity, which could be significantly enhanced by applying hole doping or pressure. Moreover, by monitoring the local spin moment and local entanglement entropy, we summarize experimentally synthesized RP nickelates into two phase diagrams. At last, we predict that a hybrid RP nickelate stacking in a ``bilayer-trilayer'' sequence may be a promising superconductor candidate.

\section{Method of calculations}
A half unit-cell crystal structure of film 2222-La327 is used to perform calculations in this work, which comes from Ref.~\cite{yue2025correlatedelectronicstructuresunconventional}. The crystal structures of 1212-La327, La$_2$NiO$_4$, and 2323-La7517 in this work were optimized by using the QUANTUM-ESPRESSO package~\cite{Giannozzi_2009}. The DFT parts of our DFT+DMFT calculations are performed by the WIEN2K code with the full-potential linearized augmented plane-wave method~\cite{Blaha-JCP152}. The generalized gradient approximation with the Perdew-Burke-Ernzerhof functional is chosen as the exchange and correlation potential~\cite{Perdew-PRL77}. The EDMFTF software package is used to perform the charge fully self-consistent DFT+DMFT calculations~\cite{Haule-PRB81}. Within about 40 DFT+DMFT cycles, we obtain a good fully self-consistent convergence. The systems are enforced to be paramagnetic. Only the Ni-$e_g$ orbitals are treated as correlated except Sr$_3$Ni$_2$O$_5$Cl$_2$, where all five Ni-3$d$ orbitals are treated as correlated. Different numbers of impurity problems are considered for different nickelates, which depend on the number of non-equivalent Ni atoms. The Hund's exchange parameter $J_H$ is set to be 1.0 eV. We choose the Coulomb interaction parameter $U$ = 8.0 eV for the Ni atoms in La$_2$NiO$_4$ and the monolayer of 1212-La5311, while a value of $U$ = 5 eV is set in the other kinds of Ni. The choices of these parameters are consistent with our previous works~\cite{PhysRevB.111.125111,PhysRevB.109.115114,PhysRevB.109.165140}, which are reasonable in the DFT+DMFT calculations of RP nickelates. The density-density form of the Coulomb repulsion is used. The finite-temperature quantum impurity problems for the DMFT are solved by the hybridization expansion continuous-time quantum Monte Carlo impurity solver~\cite{Haule-PRB75} at 200 K with an exact double-counting scheme~\cite{Haule-PRL115} for the self-energy function. Utilizing the maximum entropy method analytical continuation~\cite{Jarrell-PR269}, we obtain the real-frequency self-energy function, which is used to calculate the momentum-resolved spectral function and the other related physical quantities.

\section{Results}

\begin{figure*}[bth]
\centering
\includegraphics[width=17.2cm]{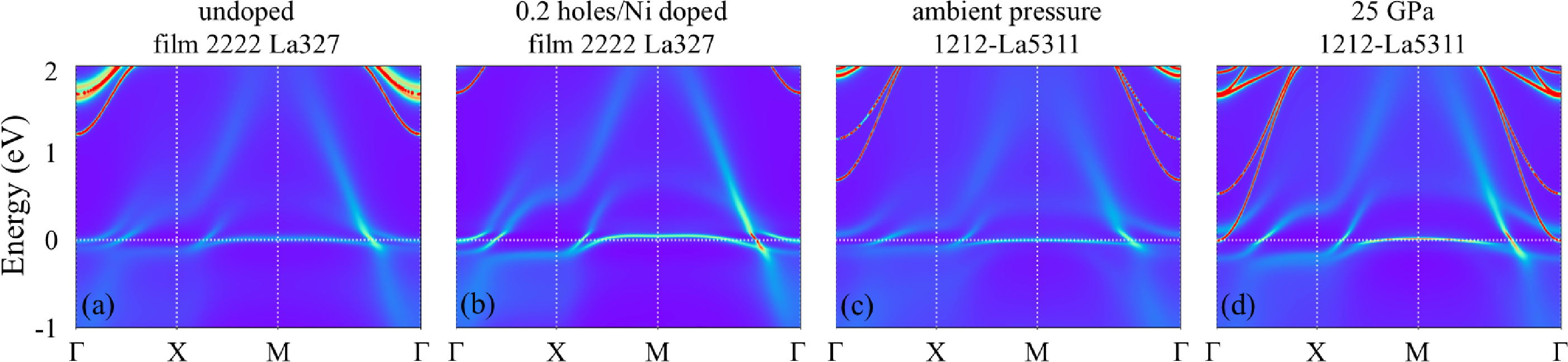}
\caption{DFT+DMFT calculated correlated electronic structures. (a) undoped film 2222-La327, (b) 0.2 holes/Ni doped film 2222-La327, (c) 1212-La5311 under ambient pressure, and (d) 1212-La5311 under 25 GPa pressure. The Hund’s exchange parameter $J_H$ is set to be 1.0 eV. And the Coulomb interaction parameter $U$ is set to be 8.0 eV for the Ni atoms in La$_2$NiO$_4$ and the monolayer of 1212-La5311, while a value of U = 5 eV is set for the other kinds of Ni. The subsequent results are calculated with the same settings.} 
\label{fig:1}
\end{figure*}

\subsection{Enhanced quasi-particle weight by hole doping in film 2222-La327}
According to the experimental results~\cite{Ko2024,Zhou2025}, superconductivity with transition temperature ($T_c$) $\sim$ 40 K was observed in the film 2222-La327 growing on a substrate of SrLaAlO$_4$ under ambient pressure. The substrate applies a $\sim$ 2\% in-plane strain for the film. No trace of superconductivity was found in other substrates with smaller in-plane strain. The angle-resolved photoemission spectroscopy (ARPES) results~\cite{li2025photoemissionevidencemultiorbitalholedoping} suggest that the superconducting state of interfaces is hole-doped. The diffusion of interfacial Sr atoms from the substrate to the surface of film was found in 3-unit-cell (3UC) thick samples of tetragonal La$_{2.85}$Pr$_{0.15}$Ni$_2$O$_7$. And measurements of resistance show that the first UC near the interface is highly conductive while the third UC is almost insulating.

In Figs.~\ref{fig:1}(a) and (b), we exhibit the fully self-consistent DFT+DMFT calculated correlated electronic structures of film 2222-La327 for both undoped (not superconducting) and 0.2 holes/Ni doped (superconducting) cases under ambient pressure. The blurry spectrum and renormalization of band width (see DFT results in Fig.~\ref{fig:6} in the Appendix A) indicate that film 2222-La327 is correlated. In both the undoped and doped cases, the $\gamma$ bands, mainly contributed by the Ni-3$d_{z^2}$ orbitals at the $M$ point, touch the Fermi level, which is similar to that of high-pressure bulk 2222-La327 except for the emergence of the $\epsilon$ pocket around the $\Gamma$ point. However, in the doped case, the quasi-particle weights of those bands around the Fermi level is enhanced, and the $\gamma$ band crosses the Fermi level. Both our results and the ARPES results~\cite{li2025photoemissionevidencemultiorbitalholedoping} suggest that the hole-doped superconducting interfaces possess enhanced quasi-particle weights, while the Fermi surfaces of the undoped parts with low quasi-particle weights are blurry.

\subsection{Enhanced quasi-particle weight by applying pressure in 1212-La5311}
\begin{figure}[h]
\centering
\includegraphics[width=8.6cm]{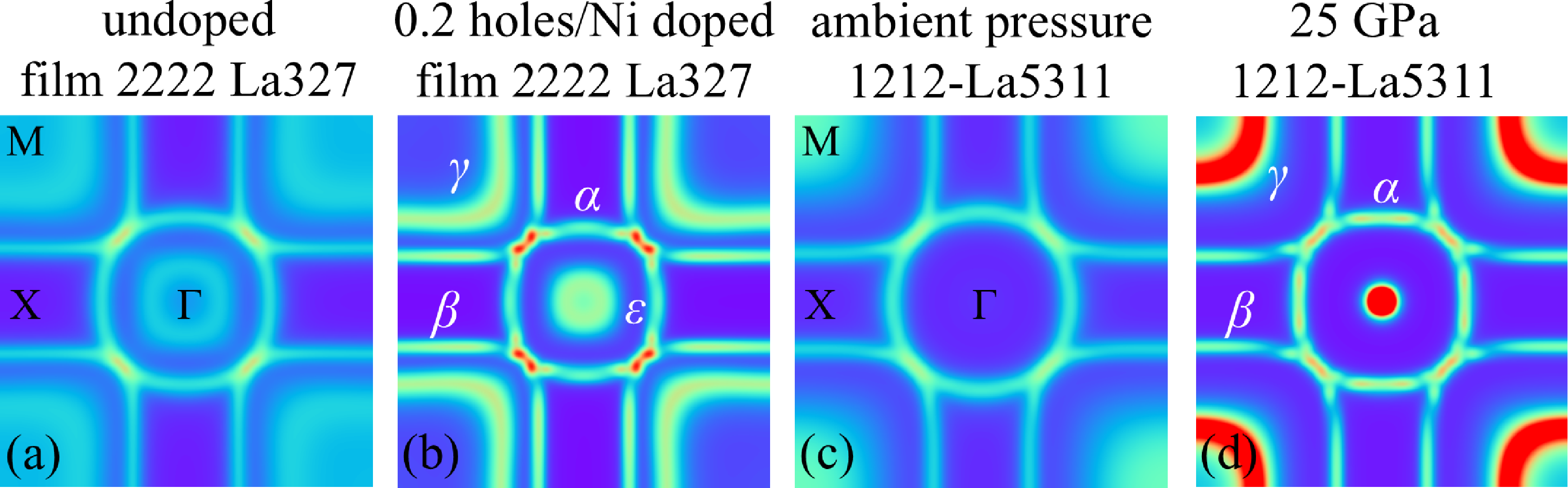}
\caption{DFT+DMFT calculated Fermi surfaces at the $k_z$ = 0 plane. (a) undoped film 2222-La327, (b) 0.2 holes/Ni doped film 2222-La327, (c) 1212-La5311 under ambient pressure, and (d) 1212-La5311 under 25 GPa pressure.}
\label{fig:2}
\end{figure}

We perform a systematic investigation regarding the correlated electronic structure of 1212-La5311 under different pressures. As shown in Figs.~\ref{fig:1}(c) and (d), the flat bands at the Fermi level are mainly contributed by the bonding state of bilayer (BL) Ni-3$d_{z^2}$ orbitals (see orbital-resolved spectral functions in Fig.~\ref{fig:9} in the Appendix B), and the blurry anti-bonding state is located at $\sim$ 0.4 eV above the Fermi level. As pressure increases, the bandwidth and the split between the bonding and anti-bonding states become larger. In Figs.~\ref{fig:2}(c) and (d), the calculated Fermi surfaces exhibit similar Fermi pockets to those of film 2222-La327. The quasi-particle weight of the $\gamma$ pocket shows significant enhancement with the increasing pressure, which suggests that pressure enhances the coherence of the Ni-3$d_{z^2}$ orbitals. On the contrary, the quasi-particle weights of $\alpha$ and $\beta$ pockets contributed by the Ni-3$d_{x^2-y^2}$ are not sensitive to pressure. Similar orbital-selective electronic correlation was also found in other high-pressure RP nickelates~\cite{PhysRevB.111.125111,PhysRevB.109.115114,PhysRevB.109.165140}. More importantly, the effect on correlated electronic structure brought by applying pressure, namely enhanced quasi-particle weight, is akin to that brought by hole doping. By defining quasi-particle weight $Z$ as 
\begin{equation}
{Z} = \bigg[  1 - \frac{\partial \mbox{Re} \Sigma (\omega)}{\partial \omega} \bigg|_{\omega = 0} \bigg]^{-1},
\label{eq:weight}
\end{equation}
we further quantify the quasi-particle weights of the Ni-$e_g$ orbital under different situations and list them in Table~\ref{tab1}. The enhancement of quasi-particle weights is found after applying hole doping or pressure.

\begin{table}[th]
\begin{center}
\small
\renewcommand\arraystretch{1.5}
\caption{Quasi-particle weights $Z$ of the Ni-$e_g$ orbtials of different cases of film 2222-La327 and the BL of 1212-La5311.}
\label{tab1}
\begin{tabular*}{\linewidth}{@{\extracolsep{\fill}} ccc}
\hline\hline
{}    & 3$d_{z^2}$  &3$d_{x^2-y^2}$  \\
\hline
film 2222-La327 @ AP  &0.167  &0.258  \\
\hline
doped film 2222-La327 @ AP  &0.254  &0.276      \\
\hline
BL of 1212-La5311 @ AP  &0.129   &0.220     \\
\hline
BL of 1212-La5311 @ 20 GPa  &0.166  &0.339      \\
\hline
BL of 1212-La5311 @ 25 GPa  &0.176  & 0.364    \\
\hline
BL of 1212-La5311 @ 30 GPa  &0.186  &0.383      \\
\hline\hline
\end{tabular*}
\end{center}
\end{table}

\subsection{Absence of superconductivity in doped La$_2$NiO$_4$ under pressure}
\begin{figure}[thb]
\centering
\includegraphics[width=8.6cm]{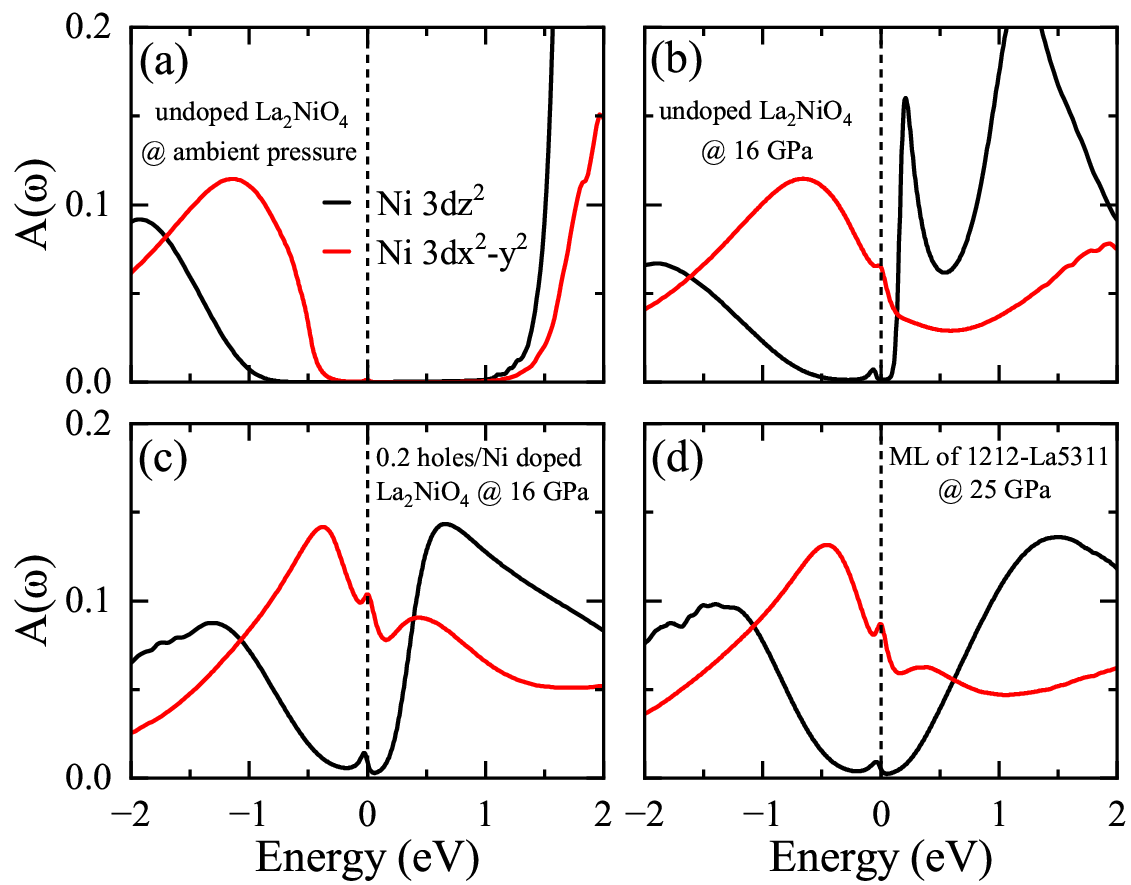}
\caption{DFT+DMFT calculated local spectral functions $A(\omega)$ of the Ni-$e_g$ orbitals. (a) undoped La$_2$NiO$_4$ under ambient pressure, (b) undoped La$_2$NiO$_4$ under 16 GPa pressure, (c) 0.2 holes/Ni doped La$_2$NiO$_4$ under 16 GPa pressure, and (d) ML of 1212-La5311 under 25 GPa pressure.}
\label{fig:3}
\end{figure}

We investigate the correlated electronic structures of La$_2$NiO$_4$ in different cases, which is also helpful in understanding the role of the monolayer of superconducting hybrid nickelates. It is known that the undoped La$_2$NiO$_4$ is an AFM Mott insulator with a Ni-3$d^8$ configuration, which was once expected to be a parent of high-$T_c$ superconductor~\cite{Rao1984}. However, an AFM metal phase was observed in hole-doped La$_2$NiO$_4$ instead of superconductivity~\cite{PhysRevB.43.1229,PhysRevLett.71.2461,PhysRevB.49.7088}. As shown in Fig.~\ref{fig:3}(a), our calculated local spectral function $A(\omega)$ of undoped La$_2$NiO$_4$ exhibits an obvious Mott gap. Under a high pressure of 16 GPa, both the Ni-$e_g$ orbitals of undoped and 0.2 holes/Ni doped cases show an orbital-selective Mott phase (OSMP) [Figs.~\ref{fig:3}(b) and (c)]. And the Mott gap becomes smaller than that of the ambient case. It is worth noting that the discovery of high-pressure superconducting RP nickelates makes people again investigate the possibility of superconductivity in La$_2$NiO$_4$ under pressure. However, La$_2$NiO$_4$ shows no trace of superconductivity under high pressures and even after applying hole doping~\cite{ZHANG2024147}. Hence, both our results and experimental reports suggest that the correlated electronic structure of OSMP with low quasi-particle weights may not favor superconductivity.

It should be noted that the monolayer of hybrid RP nickelates possesses the same chemical formula as La$_2$NiO$_4$. As shown in Fig. 3(d), the local spectral function $A(\omega)$ of the monolayer of 1212-La5311 under high pressure shows high similarities to that of hole doped La$_2$NiO$_4$ under high pressure. A very similar OSMP is observed in the monolayer (ML) Ni-$e_g$ orbitals of 1212-La5311. The ML Ni-3$d_{z^2}$ orbital shows a Mott behavior, while the ML Ni-3$d_{x^2-y^2}$ orbital is metallic, which suggests that there is charge transfer in hybrid RP nickelates and the ML Ni-$e_g$ orbitals are hole doped. Furthermore, high pressure causes a Lifshitz transition. Figure~\ref{fig:1}(d) exhibits that an La-related band at the $\Gamma$ point crosses the Fermi level under high pressure. A small hole-type pocket emerges around the $\Gamma$ point, as observed in Fig.~\ref{fig:2}(d). This phenomenon has also been found and discussed in 1313-La327~\cite{PhysRevB.111.125111}. These results convey a message that the superconductivity of hybrid RP nickelates may not come from the monolayers with OSMP.

\begin{table}[h]
\begin{center}
\small
\renewcommand\arraystretch{1.5}
\caption{DFT+DMFT calculated weights (\%) of the Ni-$e_g$ orbital local multiplets for undoped La$_2$NiO$_4$ under ambient pressure, monolayer (ML) and bilayer (BL)of 1212-La5311 under 25 GPa pressure, and 0.2 holes/Ni doped film 2222-La327 under ambient pressure, respectively. The good quantum numbers $N_{\Gamma}$ and $S_z$ denote the total occupancy and total spin of the Ni-$e_g$ orbitals, which are used to label different local spin states.}
\label{tab2}
\begin{tabular*}{\linewidth}{@{\extracolsep{\fill}} ccccccc}
\hline\hline
{$N_{\Gamma}$}   & 0   & 1   & 2   & 2   & 3   & 4   \\
\hline
{$S_z$}  & 0    & 1/2   & 0   & 1   &1/2   & 0  \\
\hline
\makecell{La$_2$NiO$_4$}  & 0.00 &1.25 & 0.92  & 79.07 & 18.29  & 0.47  \\
\hline
\makecell{ML of 1212}  & 0.02  & 4.09  & 6.42  & 67.29  & 21.42  & 0.76  \\
\hline
\makecell{BL of 1212}  & 0.32  &11.44  &22.63  &34.92  &28.40  & 2.28 \\
\hline
\makecell{film 2222}  & 0.50  &14.51  & 24.05  & 34.40  & 24.81 & 1.73  \\
\hline\hline
\end{tabular*}
\end{center}
\end{table}

\subsection{Spin fluctuation in RP nickelates}
Magnetism is a factor in unconventional superconductivity that can not be ignored. Many DFT calculations including long-range magnetic orders tried to determine magnetic ground state of RP nickelates~\cite{PhysRevB.110.205122,PhysRevB.108.L180510,PhysRevB.110.L140508}. The magnetic ground state of bulk 2222-La327 under ambient pressure is found to be a double spin-charge stripe order~\cite{Chen2024rixs,plokhikh2025unravelingspindensitywave}. 

Here, our DFT+DMFT calculations give the weights of local spin multiplets describing local spin fluctuation. As shown in Table~\ref{tab2}, the local spin multiplets of the Ni-$e_g$ orbitals of the undoped La$_2$NiO$_4$ favor a dominating state of half-filling of the $e_g$ orbtials with $S_z$ = 1~\cite{PhysRevResearch.5.033113}. The ML Ni-$e_g$ orbitals of 1212-La5311 under 25 GPa pressure exbibit similar local spin multiplets with a dominating high spin state $S_z$ = 1. These also confirm that there is a high similarity between La$_2$NiO$_4$ and the ML of 1212-La5311. As for the Ni-$e_g$ orbitals of the bilayer of 1212-La5311 and film 2222-La327, the spin fluctuation is significantly enhanced. The high spin state $S_z$ = 1 is no longer dominating and only wins the competition by a narrow margin, which suggests Hundness in RP nickelates~\cite{PhysRevB.111.125111,PhysRevB.109.115114,PhysRevB.109.165140}. Moreover, we find that the other known superconducting RP nickelates all show strong local spin fluctuation (see in Table~\ref{tab3} in the Appendix D), which suggests the importance of spin fluctuation to the superconductivity of RP nickelates.

\begin{figure}[htb]
\centering
\includegraphics[width=8.2cm]{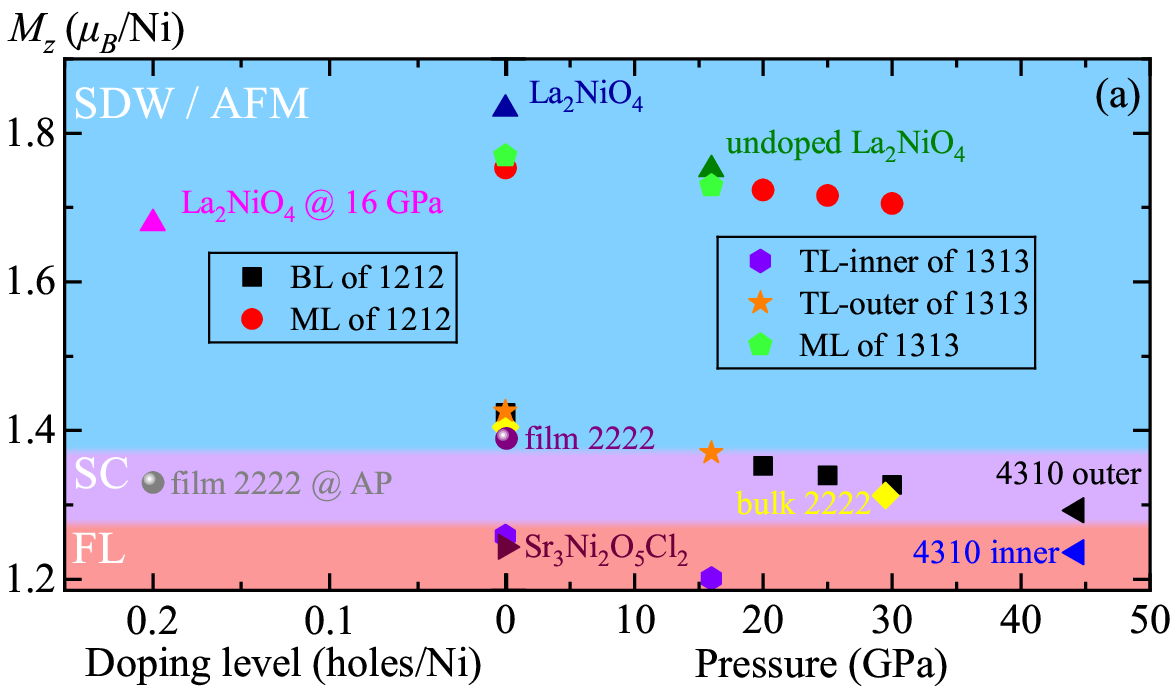}
\includegraphics[width=8.2cm]{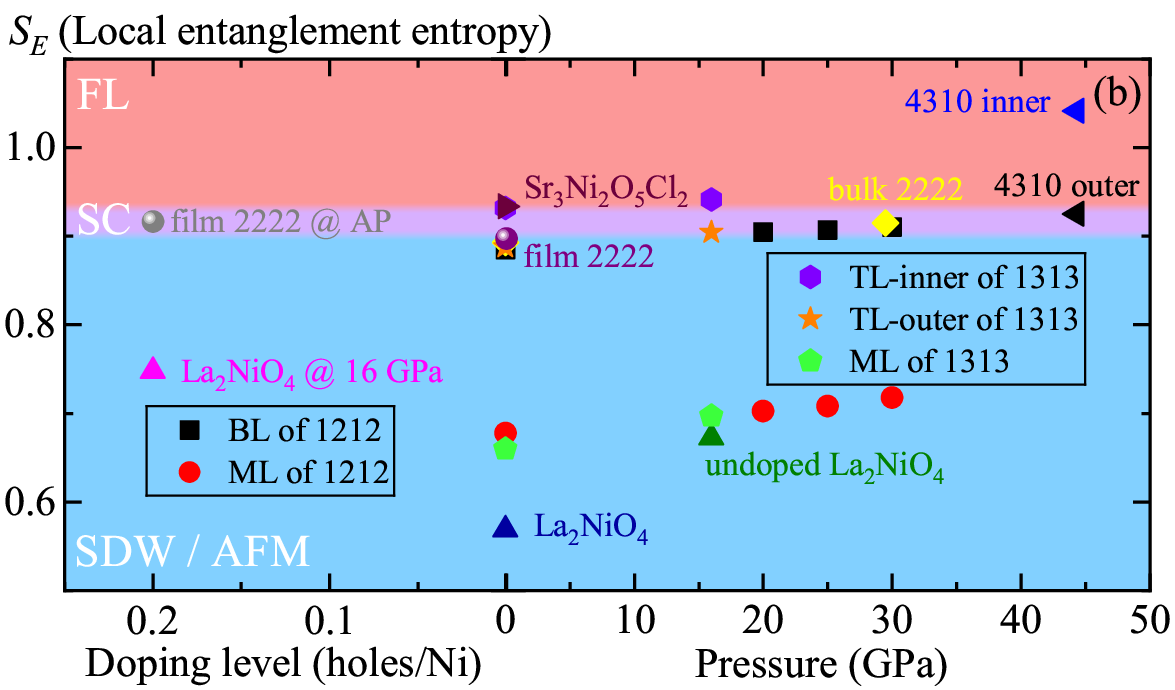}
\caption{Phase diagrams of RP nickelates with local spin moment and local entanglement entropy. The local spin moments (a) and local entanglement entropy (b) of nickelates are calculated from the local spin multiplets states of the Ni-$e_g$ orbitals, which were obtained from DFT+DMFT calculations The results of bulk 2222-La327, La4310, and hybrid 1313-La327 come from Refs.~\cite{PhysRevB.111.125111,PhysRevB.109.115114,PhysRevB.109.165140}.}
\label{fig:4}
\end{figure}

\subsection{Phase diagrams of RP nickelates with local spin moment and local entanglement entropy}
Furthermore, we define a local spin moment of Ni as $M_z$ = $g\sqrt{\langle{S_z^2}\rangle}$ ($g$ is set to 2 for the 3$d$ orbtials) and summarize the known RP nickelates into a phase diagram including magnetism, superconductivity, and Fermi liquid in Fig.~\ref{fig:4}(a). At the top side of the phase diagram, magnetism is dominant. The undoped La$_2$NiO$_4$ under ambient pressure exhibits the largest $M_z$ and shows an AFM Mott insulating behavior. Applying hole doping or pressure could reduce $M_z$ being close to that of doped or pressurized La$_2$NiO$_4$. The RP nickelates under ambient pressure, including undoped bulk and film 2222-La327, 1212-La5311, and 1313-La327, possess a smaller $M_z$ $\sim$ 1.4 $\mu_B$/Ni. Absence of superconductivity but a likely SDW transition is their common feature. These systems may locate in a region with fierce competition between magnetism and superconductivity. Both the previous~\cite{PhysRevB.111.125111,PhysRevB.109.115114,PhysRevB.109.165140} and here calculated correlated electronic structures and local spin multiplets of those superconducting RP nickelates strongly confirm that applying hole doping or pressure could enhance the spin fluctuation as well as the quasi particles, where the former provides suitable AFM correlation as glue to the latter for unconventional superconducting pairing~\cite{PhysRevLett.131.236002,PhysRevB.109.165154,PhysRevLett.132.146002,PhysRevB.111.035108,PhysRevB.108.L140504}. A model study also suggested that hole doping at low pressures may achieve a similar effect to high pressures and weaken spin correlations, as well as potentially suppress the possible SDW and favor superconductivity~\cite{PhysRevB.110.235119}.

However, it should be noted that there is a delicate balance and over doping may be harmful to superconductivity. For example, Sr$_3$Ni$_2$O$_5$Cl$_2$ with a nominal Ni-3$d^7$ configuration was theoretically predicted to be superconducting~\cite{PhysRevB.111.064511}, while a later experimental report found no superconductivity~\cite{yamane2024highpressuresynthesisbilayernickelate}. Our calculations suggest that both of the correlated electronic structure (see in Fig.~\ref{fig:10} in the Appendix C) and $M_z$ of Sr$_3$Ni$_2$O$_5$Cl$_2$ are very close to those of the inner Ni of La4310~\cite{PhysRevB.109.165140} as well as the trilayer in 1313-La327~\cite{PhysRevB.111.125111}, which show a Fermi-liquid behavior. This may be explained by the filling of the Ni-$e_g$ orbitals. Ni has a valence of +3 in Sr$_3$Ni$_2$O$_5$Cl$_2$, which is 0.25 hole doped per Ni-$e_g$ orbital compared to the Ni$^{2.5+}$ case of 2222-La327 and is 0.5 hole doped compared to the half-filling case of Ni$^{2+}$. This in turn supports that the weakly correlated behaviors for the Ni-$e_g$ orbitals of the inner-layer Ni in La4310~\cite{PhysRevB.109.165140} and the inner-layer Ni of the TL in 1313-La327~\cite{PhysRevB.111.125111} may be accounted for by over doping and an approximate +3 valence of the Ni there. These examples suggest that the conditions for realizing superconductivity are strict and over doping may make AFM correlation too weak to support an unconventional pairing, just like the case of over doped cuprates.

In Fig.~\ref{fig:4}(b), we show a phase diagram of RP nickelates with local entanglement entropy $S_E$ as a parameter, where $S_E$ = $-\Sigma_{(N_\Gamma, S_z)}$ ${\rho_{(N_\Gamma, S_z)}}\log{\rho_{(N_\Gamma, S_z)}}$ describes the degree to which the electrons in the $e_g$ orbitals of a Ni atom are entangled with external electrons and $\rho_{(N_{\Gamma}, S_z)}$ is the probability of local spin multiplets exhibited in Table~\ref{tab2} and~\ref{tab3}. The phase diagram regarding the $S_E$ also suggests that superconductivity is located in a narrow region between local states (magnetism) and itinerant states (Fermi liquid).

\begin{figure}[tbh]
\centering
\includegraphics[width=8cm]{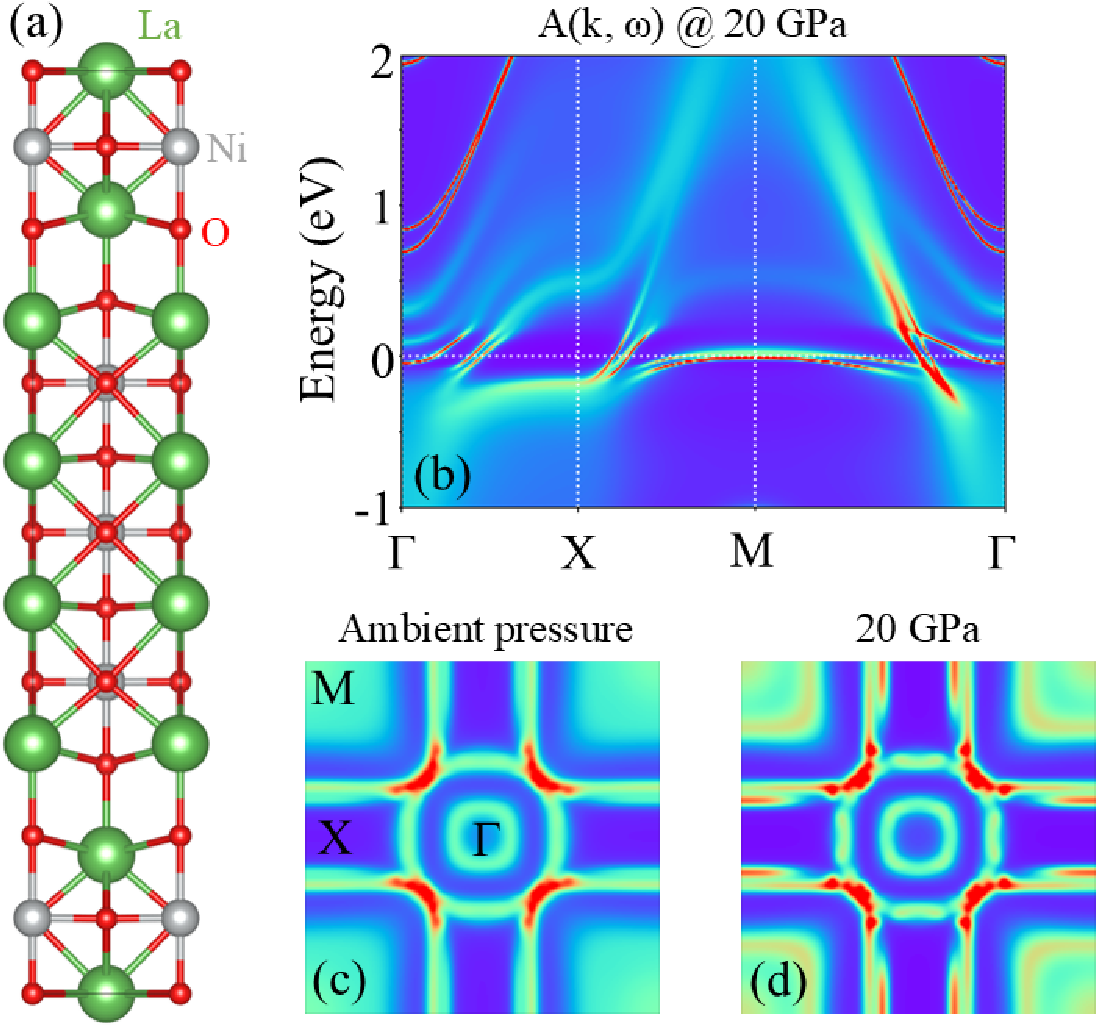}
\caption{Results of tetragonal 2323-La7517. (a) crystal structure. (b) The DFT+DMFT calculated correlated electronic structures under 20 GPa pressure. The DFT+DMFT calculated Fermi surfaces at $k_z$ = 0 plane under (c) ambient pressure and (d) 20 GPa pressure.}
\label{fig:5}
\end{figure}

\subsection{Possible superconductivity in a hybrid bilayer-trilayer RP nickelate}
Following the above results and analysis, we propose a possible superconducting RP nickelate, which holds a stacking of  ``bilayer-trilayer'' sequence, namely La$_7$Ni$_5$O$_{17}$ (2323-La7517), as shown in Fig.~\ref{fig:5}(a). Our DFT+DMFT calculations suggest that the bonding states of both bilayer and trilayer at the $M$ point emerge around the Fermi level and show large quasi-particle weights [Figs.~\ref{fig:5}(b-d)]. And the local spin multiplets show large spin fluctuation (see in Table~\ref{tab3} in the Appendix D). Applying pressure may cause it to fall further into the region of superconductivity in Fig.~\ref{fig:4}. Hence, it is a very promising superconductor candidate. Considering the successful synthesis of film 2222-La327~\cite{Ko2024,Zhou2025} and La4310~\cite{li2025signaturesuperconductivitypressurizedla4ni3o10x} under ambient pressure, constructing a bilayer-trilayer heterostructure would also be an interesting attempt besides the synthesis of bulk samples.

\section{Discussion and Conclusion}
The electronic structure is considered a crucial factor in the superconductivity of RP nickelates. A recent study suggested that oxygen vacancy significantly changes the electronic structure of film 2222-La327 and destroys superconductivity~\cite{wang2025}. The multiorbital nature of the superconducting electronic states and the
critical role of $z$-axis orbitals were emphasized~\cite{wang2025}, which is consistent to our studies that the quasi-particle weights of electrons, especially for the Ni-3$d_{z^2}$ ortbials. The enhancement of quasi-particle weights by applying hole doping or pressure is an effective means to realizing superconductivity in RP nickelates.

On the other hand, magnetism is also an important issue for unconventional superconductivity. The claim that AFM fluctuation provides glue for unconventional superconducting paring has been widely accepted in cuprates and iron pnictides. For example, the importance of Fermi surface nesting between the $\Gamma$ and $M$ points to superconductivity was once proposed for the unconventional superconductor LaOFeAs~\cite{PhysRevB.78.033111,PhysRevLett.101.057003,PhysRevLett.101.087004}.  However, the absence of hole-type pocket at the $\Gamma$ point in superconducting monolayer FeSe~\cite{PhysRevLett.102.177003,PhysRevLett.117.117001} suggests a different scenario and the dominating role of AFM fluctuation for unconventional superconductivity comes into focus~\cite{PhysRevB.78.224517,PhysRevB.98.020507}. In RP nickelates, SDW exhibits intimate relation to superconductivity. An experimental report found that tetragonal La4310 shows neither SDW nor superconductivity under both ambient and high pressures, which implies the importance of SDW to the emergence of  superconductivity~\cite{shi2025absencesuperconductivitydensitywavetransition}. Our calculations suggest that both too strong or too weak local spin moments may be poisonous to superconductivity, while strong spin fluctuation is favorable.

In summary, we perform systematical studies regarding film 2222-La327, hybrid 1212-La5311, and La$_2$NiO$_4$ by using self-consistency DFT+DMFT calculations. We find that both hole doping and applying pressure could enhance the quasi-particle weights and local spin fluctuation of the Ni-$e_g$ orbitals. A comprehensive comparison of the known RP nickelates suggests the importance of these two factors to the superconductivity. Furthermore, by monitoring the local spin moment and local entanglement entropy, we establish phase diagrams that include the experimentally synthesized RP nickelates. At last, following the two key factors, we propose that a ``bilayer-trilayer'' phase of RP nickelate may be a promising superconductor candidate.

\begin{acknowledgments}
This work was supported by the National Key R\&D Program of China (Grants No. 2024YFA1408601 and No. 2024YFA1408602) and the National Natural Science Foundation of China (Grant No. 12434009). Z.O. was also supported by the Outstanding Innovative Talents Cultivation Funded Programs 2025 of Renmin University of China. Z.Y.L. was also supported by the Innovation Program for Quantum Science and Technology (Grant No. 2021ZD0302402). Computational resources were provided by the Physical Laboratory of High Performance Computing in Renmin University of China.
\end{acknowledgments}

\newpage
\appendix
\section{DFT results}
\begin{figure}[hb]
\centering
\includegraphics[width=7.6cm]{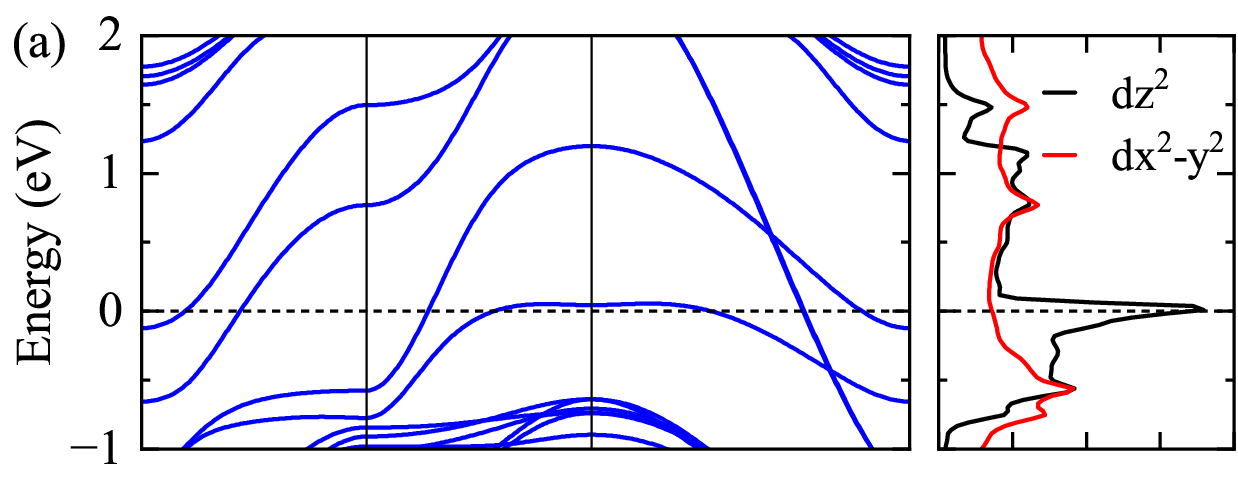}
\includegraphics[width=7.6cm]{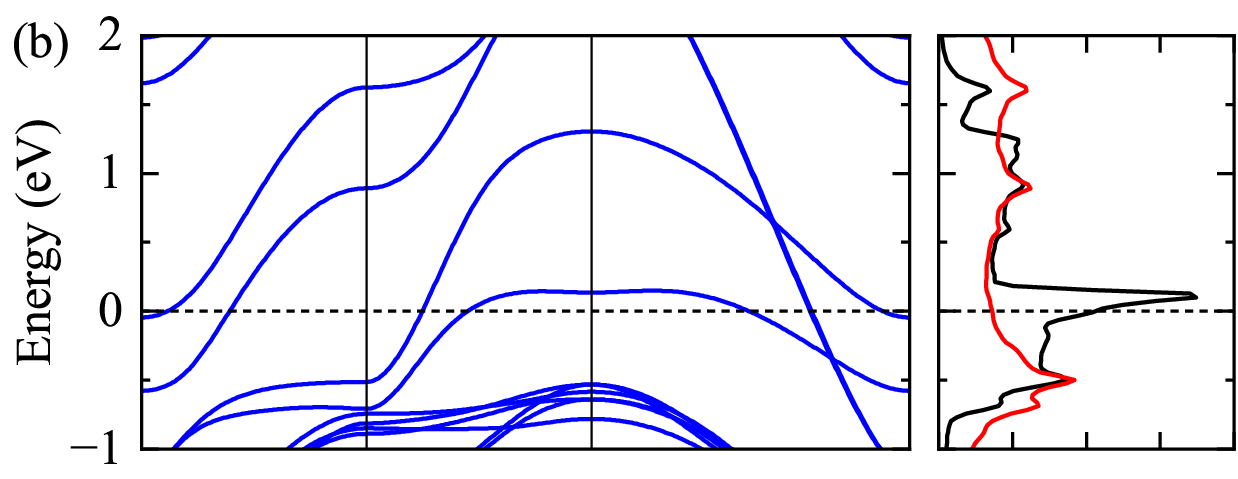}
\includegraphics[width=7.6cm]{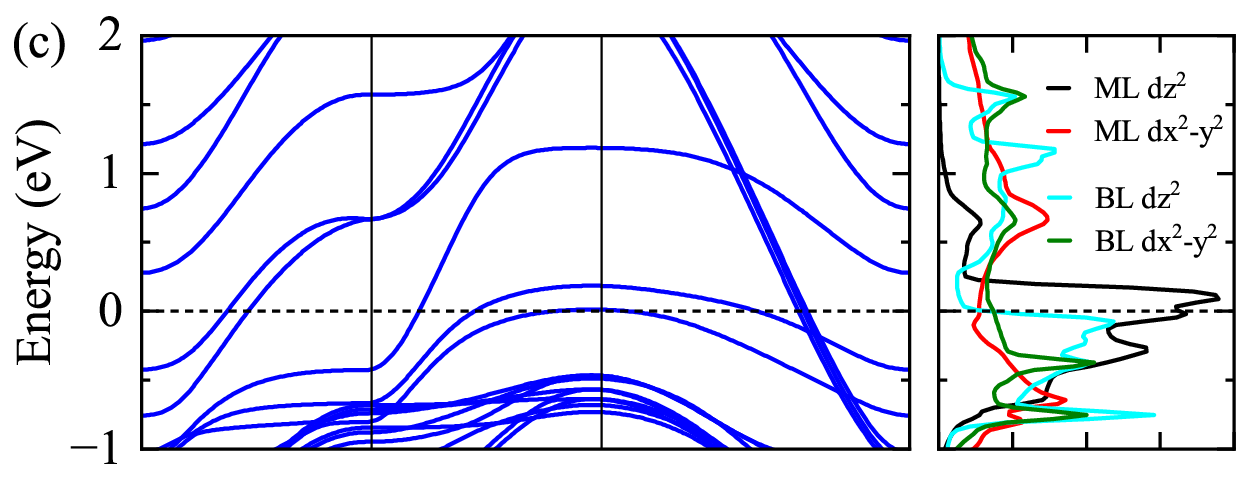}
\includegraphics[width=7.6cm]{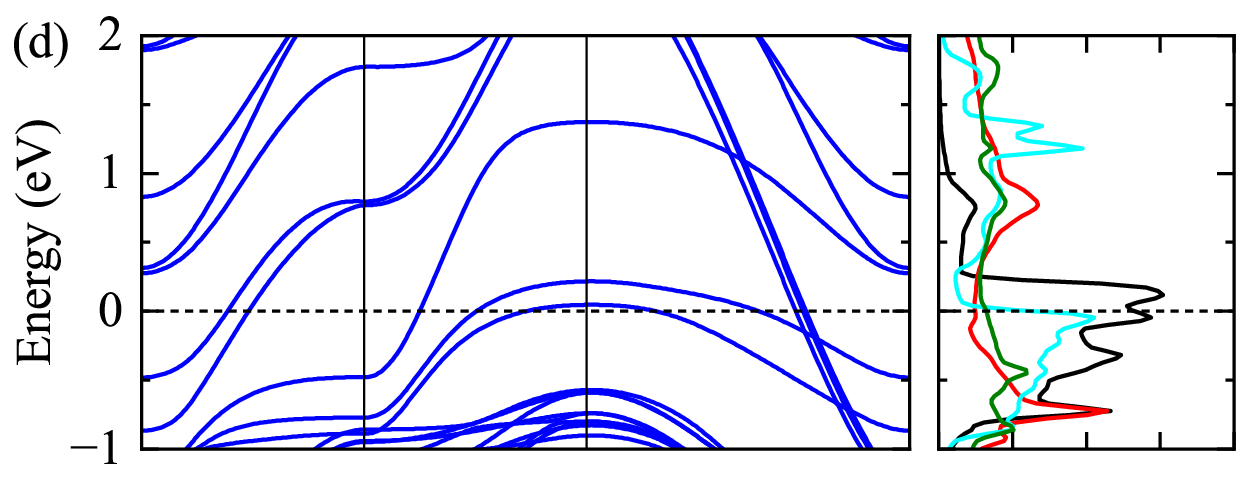}
\includegraphics[width=7.6cm]{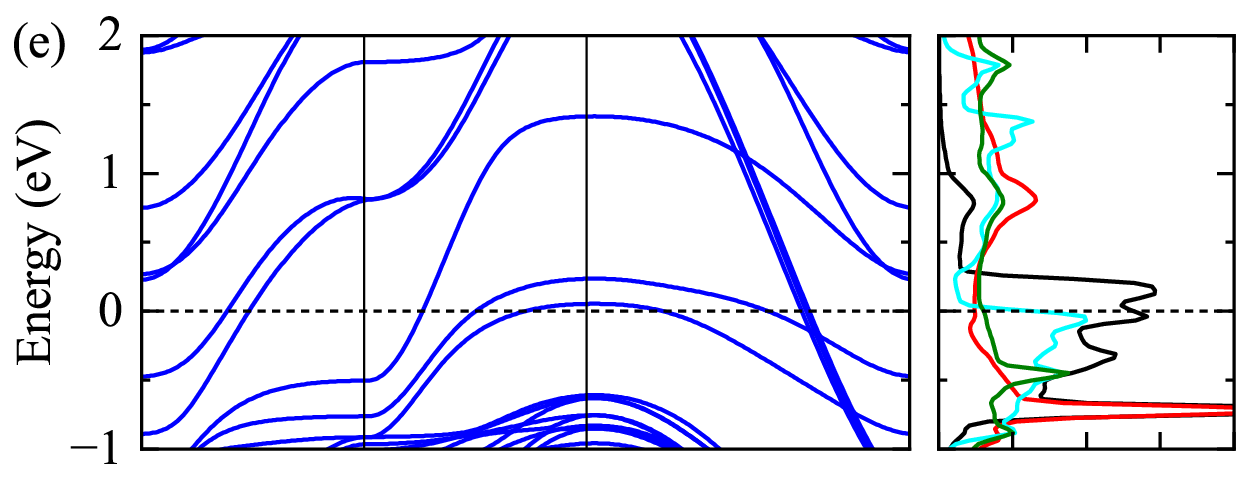}
\includegraphics[width=7.6cm]{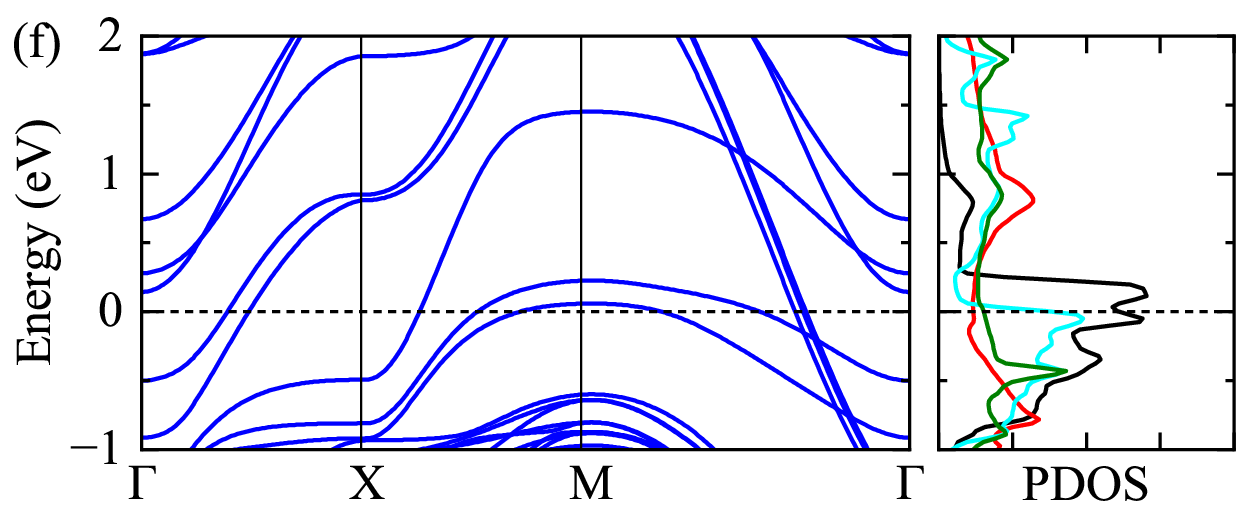}
\caption{DFT calculated band structures. (a) undoped film 2222-La327. (b) 0.2 holes/Ni doped film 2222-La327. (c-f) 1212-La5311 under ambient, 20 GPa, 25 GPa, and 30 GPa pressure, respectively. }
\label{fig:6}
\end{figure}

The DFT calculated band structures of film 2222-La327 and 1212-La5311 in Fig.~\ref{fig:6} show that the bandwidth of the Ni-$e_g$ orbitals is $\sim$ 4 eV. However, the corresponding results obtained by DFT+DMFT calculations in the main text show blurry spectrums with a smaller bandwidth. These findings suggest that these RP nickelates are strongly correlated.

\section{DFT+DMFT results of 1212-La5311}
\begin{figure}[hb]
\centering
\includegraphics[width=8.6cm]{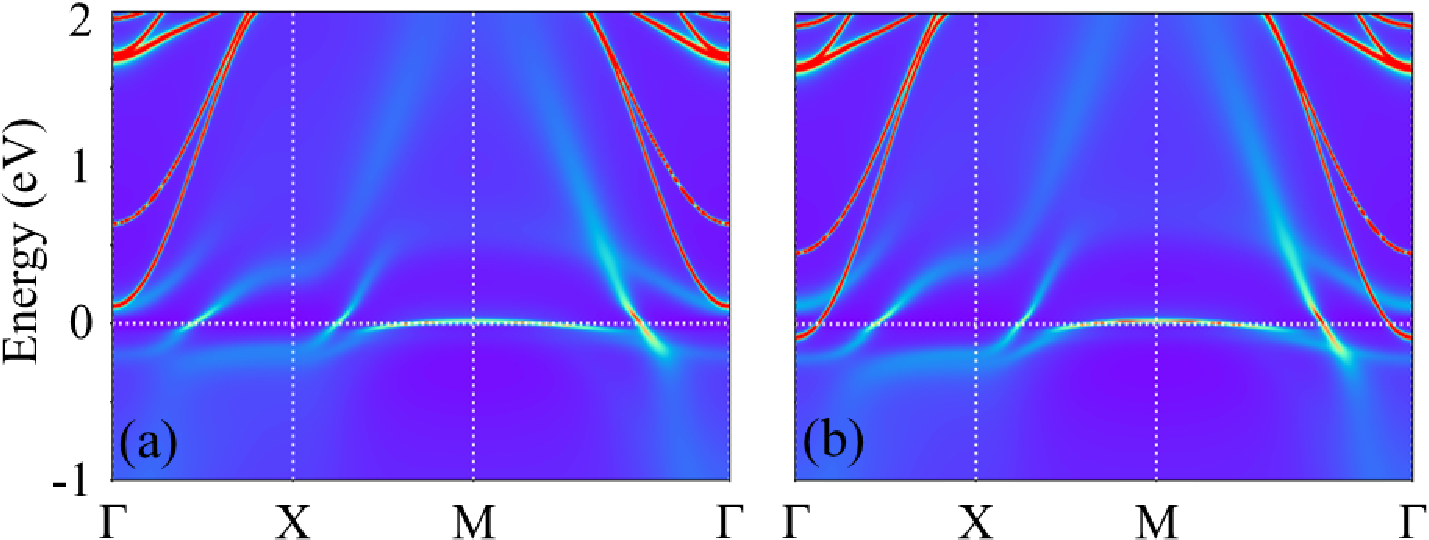}
\caption{DFT+DMFT calculated spectral functions $A(k, \omega)$ of 1212-La5311 under (a) 20 GPa pressure and (b) 30 GPa pressure, respectively.}
\label{fig:7}
\end{figure}

\begin{figure}[hb]
\centering
\includegraphics[width=6.5cm]{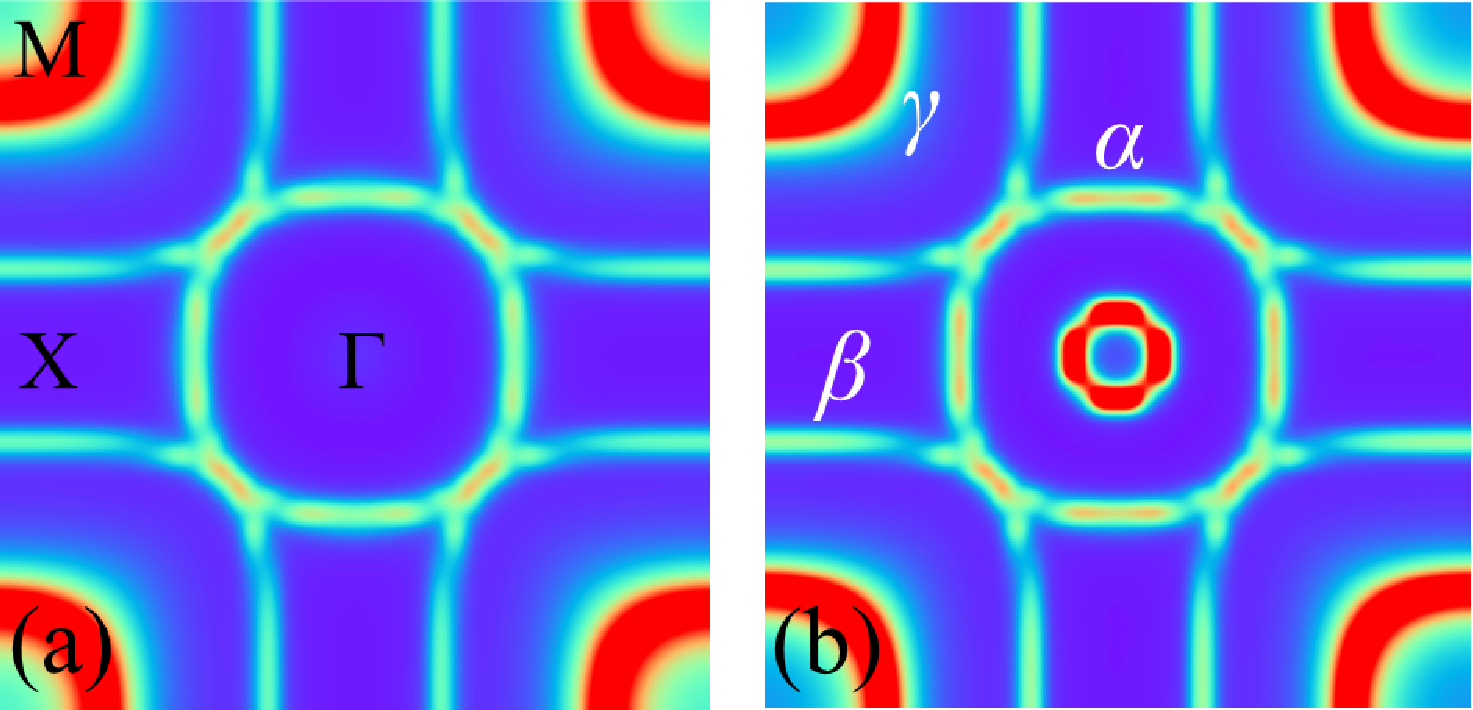}
\caption{DFT+DMFT calculated Fermi surfaces of 1212-La5311 under (a) 20 GPa pressure and (b) 30 GPa pressure, respectively.}
\label{fig:8}
\end{figure}

\begin{figure}[thb]
\centering
\includegraphics[width=8.6cm]{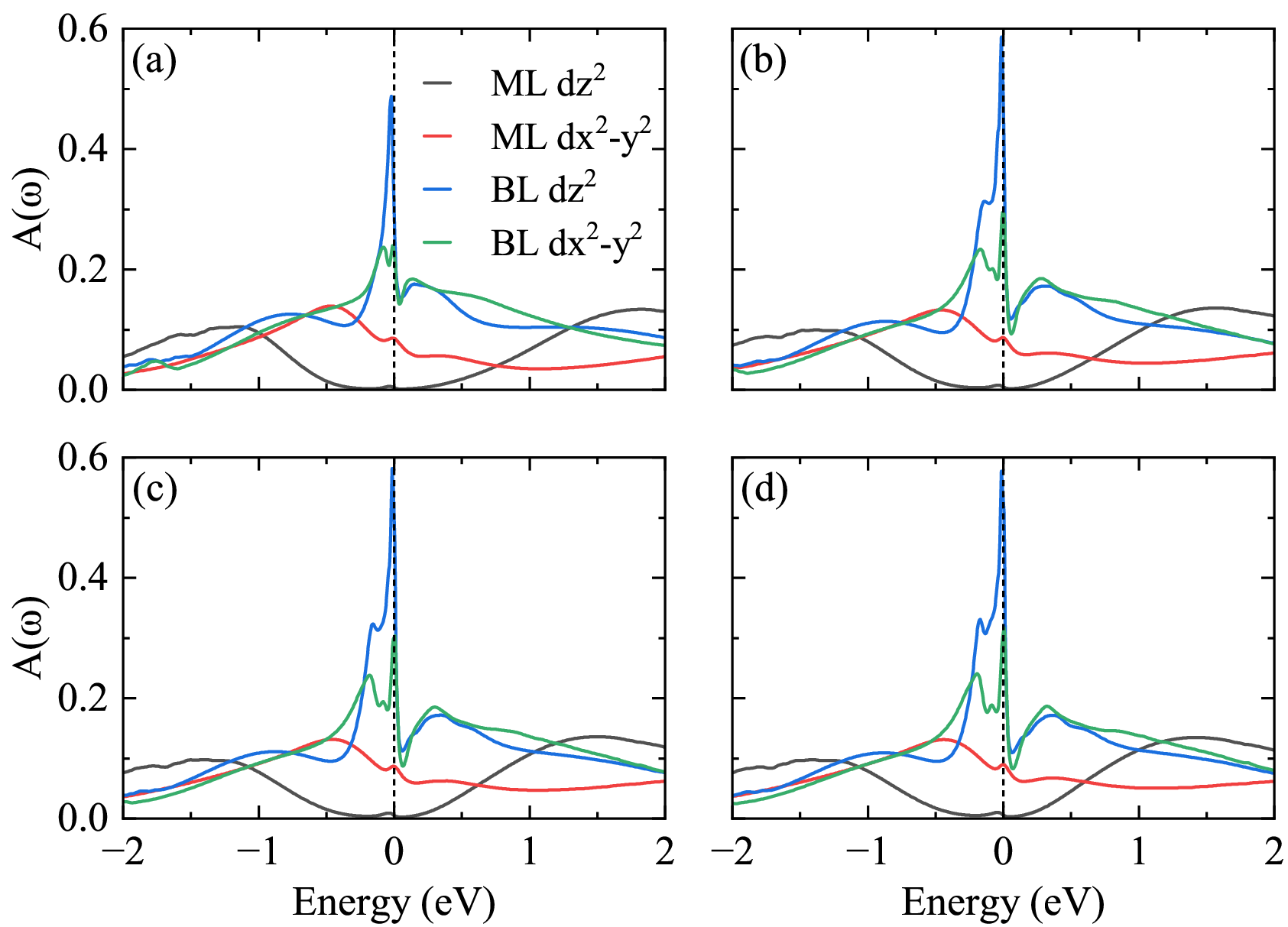}
\caption{DFT+DMFT calculated spectral functions $A(\omega)$ of the Ni-$e_g$ orbitals of 1212-La5311. (a) ambient pressure, (b) 20 GPa pressure, (c) 25 GPa pressure, and (d) 30 GPa pressure, respectively.}
\label{fig:9}
\end{figure}

Here, we show the DFT+DMFT calculated $A(k, \omega)$ of 1212-La5311 under 20 and 30 GPa pressures in Fig.~\ref{fig:7}. Combining with the results of the cases of ambient and 25 GPa pressures shown in the main text, a Lifshitz transition under high pressure is clearly exhibited. An La-related band lies above the Fermi level under ambient pressure. When increasing pressure, this band touches the Fermi level under $\sim$ 25 GPa pressure. An obvious Fermi pocket emerges around the $\Gamma$ point under a higher pressure of 30 GPa [Fig.~\ref{fig:8}(b)]. The DFT+DMFT calculated orbital-resolved spectral functions $A(\omega)$ of 1212-La5311 in Fig.~\ref{fig:9} show that the ML Ni shows an orbital-selective Mott behavior under both ambient and high pressures, which is similar to the case of hole-doped high-pressure La$_2$NiO$_4$.

\section{Results of Sr$_3$Ni$_2$O$_5$Cl$_2$}
\begin{figure}[h]
\centering
\includegraphics[width=6.8cm]{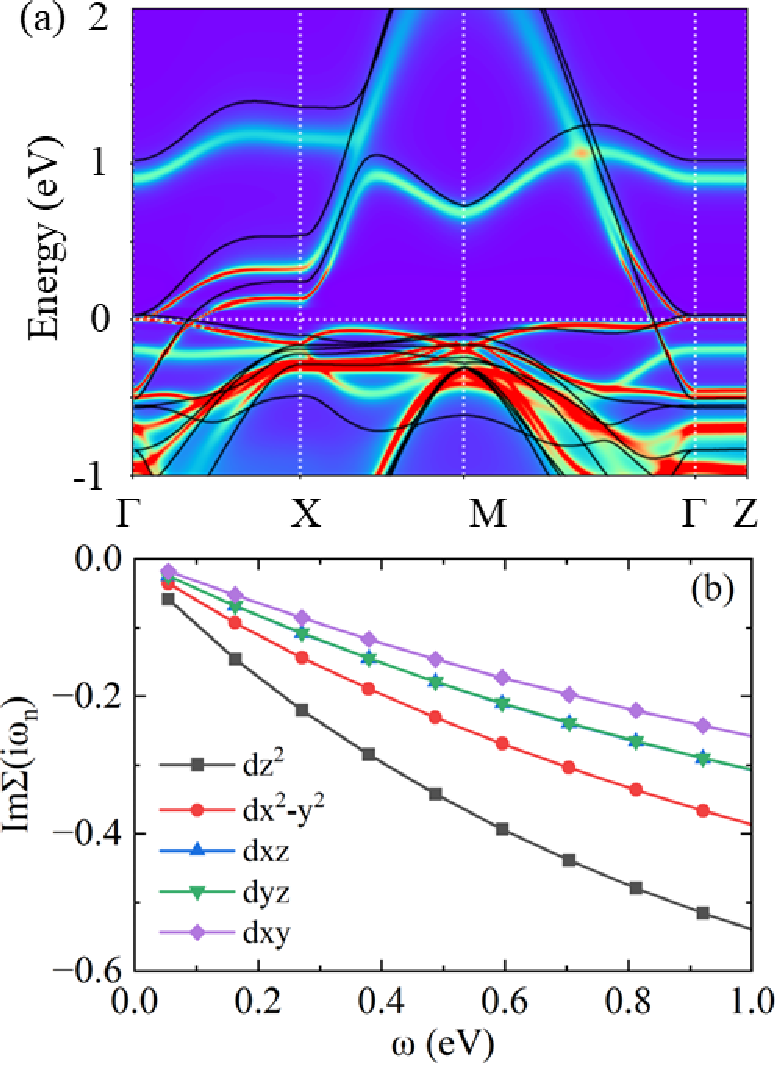}
\caption{DFT+DMFT calculated results of Sr$_3$Ni$_2$O$_5$Cl$_2$. (a) Spectral function $A(k, \omega)$. The black lines denote the DFT band structure. (b) Imaginary parts of the self-energy functions of the Ni-3$d$ orbitals. }
\label{fig:10}
\end{figure}

As shown in Fig.~\ref{fig:10}, the DFT+DMFT calculated spectral function $A(k, \omega)$ of Sr$_3$Ni$_2$O$_5$Cl$_2$ shows good coherence as well as weak band renormalization. The imaginary parts of the self-energy functions of the Ni-3$d$ orbitals show a linear behavior at low frequencies, and the corresponding Im$\Sigma$(i${\omega}_n$) are close to zero around zero frequency. All these results suggest a Fermi liquid behavior for Sr$_3$Ni$_2$O$_5$Cl$_2$.

\section{DFT+DMFT calculated local multiplets}
\begin{table*}
\begin{center}
\small
\renewcommand\arraystretch{1.5}
\caption{DFT+DMFT calculated weights (\%) of the Ni-$e_g$ orbital local multiplets of different RP nickelates. The good quantum numbers $N_\Gamma$ and $S_z$ denote the total occupancy and total spin of the Ni-$e_g$ orbitals, respectively, which are used to label different local spin states.}
\label{tab3}
\begin{tabular*}{\linewidth}{@{\extracolsep{\fill}} cccccccc}
\hline\hline
\multirow{2}{*}{P (GPa)} &{$N_{\Gamma}$}   & 0   & 1   & 2   & 2   & 3   & 4   \\
 &{$S_z$}  & 0    & 1/2   & 0   & 1   &1/2   & 0  \\
\hline
0 &undoped film 2222  & 0.30 &11.59 & 21.05 & 38.76  & 26.43  & 1.87  \\
\hline
0 &0.2 holes/Ni doped film 2222  & 0.50  & 14.51  & 24.05  & 34.40  & 24.81  & 1.73 \\
\hline
0 & \multirow{4}{*}{ML of 1212}  & 0.02  & 3.81  & 5.66  & 71.24  & 18.75  & 0.52  \\

20 &  & 0.02  & 4.06  & 6.25  & 68.03  & 20.93  & 0.71  \\

25 &  & 0.02  & 4.09  & 6.42  & 67.29  & 21.42  & 0.76  \\

30 &  & 0.03  & 4.30  & 6.87  & 66.19  & 21.82  & 0.80  \\
\hline
0 & \multirow{4}{*}{BL of 1212}  & 0.25  & 10.44  & 19.70  & 41.49  & 26.36  & 1.77  \\

20 &   & 0.31  & 11.35  & 22.24  & 35.89  & 28.03  & 2.18  \\

25 &  & 0.32  & 11.44  & 22.63  & 34.92  & 28.40  & 2.28  \\

30 &  & 0.34  & 11.64  & 23.07  & 33.93  & 28.67  & 2.36  \\
\hline
0 &\multirow{2}{*}{undoped La$_2$NiO$_4$}  & 0.00  & 1.25  & 0.92  & 79.07  & 18.29  & 0.47  \\

16 &  & 0.02  & 3.40  & 4.42  & 70.63  & 20.86  & 0.66  \\
\hline
16 &0.2 holes/Ni doped La$_2$NiO$_4$  & 0.04  & 5.89  & 9.38  & 63.98  & 20.06  & 0.65  \\
\hline
0 &Sr$_3$Ni$_2$O$_5$Cl$_2$  &0.87  & 16.83  & 26.78  & 28.09  & 25.48  & 1.94  \\
\hline
0 &\multirow{2}{*}{BL of 2323}  &0.33  & 11.78  & 21.33  & 39.14  & 25.70  & 1.72  \\

20 &   &0.39  & 12.43  & 23.13  & 34.80  & 27.17  & 2.09  \\
\hline
0 &\multirow{2}{*}{TL outer of 2323}  &0.42  & 13.01  & 23.00  & 36.50  & 25.35  & 1.73  \\

20 &  &0.49  & 13.67  & 24.38  & 32.83  & 26.58  & 2.06  \\
\hline
0 &\multirow{2}{*}{TL inner of 2323}  &0.61  & 15.32  & 25.33  & 32.55  & 24.49  & 1.70  \\

20 &  &0.81  & 16.93  & 26.56  & 29.20  & 24.62  & 1.89  \\
\hline\hline
\end{tabular*}
\end{center}
\end{table*}

\newpage
\bibliography {lno1212}

\end{document}